\documentclass[10pt,a4paper]{article}
\usepackage{amssymb,amsmath, amsfonts}
\usepackage{graphicx,graphics}
\usepackage{mathtools}
\usepackage{bbold}
\usepackage[english]{babel}
\usepackage[utf8]{inputenc}
\usepackage{epsfig,url}
\usepackage{bbm,theorem}
\usepackage{a4wide}
\usepackage{color}
\usepackage{enumerate}
\usepackage{calrsfs}
\usepackage[bookmarks=true]{hyperref}
\usepackage{bookmark}
\usepackage{amsmath}
\usepackage{amsfonts}
\usepackage{amssymb}
\usepackage{verbatim}
\usepackage{graphicx}
\usepackage[normalem]{ulem}
\providecommand{\U}[1]{\protect\rule{.1in}{.1in}}
\DeclareMathAlphabet{\pazocal}{OMS}{zplm}{m}{n}
\newtheorem{theorem}{Theorem}[section]
\newtheorem{definition}[theorem]{Definition}

\newtheorem{lemma}[theorem]{Lemma}
\newtheorem{proposition}[theorem]{Proposition}

{\theorembodyfont{\upshape}

}
\numberwithin{equation}{section}
\numberwithin{theorem}{section}
\newcommand{\qed}{\hfill$\Box$}

\newcommand{\ep}{\varepsilon}

\newcommand{\R}{\mathbb{R}}
\newcommand{\N}{\mathbb{N}}

\newcommand{\PP}{\mathbb{P}}
\newcommand{\EE}{\mathbb{E}}

\newcommand{\mbf}{\mathbf}

\def\e{\varepsilon}\newcounter{jlisti}

\title{Two-dimensional Lorentz process \\ for magnetotransport: \\ Boltzmann-Grad limit}
\author{Alessia Nota, Chiara Saffirio, Sergio Simonella}

\date{\today}

\def\adresse{
\begin{description}

\item[A. Nota:] Dipartimento di Ingegneria e Scienze dell'Informazione e Matematica, \\ Universit\`a degli studi dell'Aquila, via Vetoio, 67100 L'Aquila, Italy\\ 
E-mail: \texttt{alessia.nota@univaq.it}

\item[C. Saffirio:] Departement Mathematik und Informatik, \\ Universit\"at Basel, Spiegelgasse 1, CH-4051 Basel, Switzerland\\
E-mail: \texttt{chiara.saffirio@unibas.ch}

\item[S. Simonella:] CNRS and UMPA (UMR CNRS 5669) \\
 \'{E}cole Normale Sup\'{e}rieure, 46 all\'ee d'Italie,
69364 Lyon,
France\\
E-mail: \texttt{sergio.simonella@ens-lyon.fr}
\end{description}
}

\begin{document}
\maketitle

\abstract{\noindent We study a system of charged, noninteracting classical particles moving in a Poisson distribution of hard-disk scatterers in two dimensions, under the effect of a magnetic field perpendicular to the plane. We prove that, in the low-density (Boltzmann-Grad) limit, the particle distribution evolves according to a generalized linear Boltzmann equation, previously derived and solved by Bobylev et al.\;\cite{BMHH, BMHH1, BHPH}. In this model, Boltzmann's chaos fails, and the kinetic equation includes non-Markovian terms. 
The ideas of  \cite{Ga69} can be however adapted to prove convergence of the process with memory. 
\bigskip

\begin{center}
{\bf R\'esum\'e}
\end{center}

\noindent Nous \'etudions un syst\`eme de particules classiques charg\'ees, qui n'interagissent pas,  se d\'epla\c{c}ant dans une distribution de Poisson de disques durs en deux dimensions, sous l'effet d'un champ magn\'etique perpendiculaire au plan. Nous montrons que, dans la limite de densit\'e faible (Boltzmann-Grad), la distribution des particules \'evolue selon une \'equation  de Boltzmann lin\'eaire g\'en\'eralis\'ee, d\'ej\`a obtenue et r\'esolue par Bobylev et al.\;\cite{BMHH, BMHH1, BHPH}. Dans ce mod\`ele, le chaos de Boltzmann n'est pas v\'erifi\'e et l'\'equation cin\'etique inclut des termes non markoviens.
Les id\'ees de \cite{Ga69} peuvent cependant \^etre adapt\'ees pour prouver la convergence du processus avec m\'emoire.}

\bigskip
\noindent
{\it Keywords:} Lorentz gas; magnetic field; generalized Boltzmann equation; low-density limit; non-Markovian process; memory terms.

\bigskip

\section{Introduction}

We consider a uniform  Poisson distribution of hard disks (scatterers, obstacles) of radius $\ep>0$ in $\R^2$ and denote by { $c_1,c_2,\dots \in
\R^2$} their centers. 
Given $\mu>0$, the probability of finding $N$ obstacles in a bounded measurable set $\mathcal{A}\subset\R^2$ and with positions $c_1,\dots,c_N$
is 
\begin{equation}\label{poisson}
\PP_{N,\mathcal{A}}(\,d\mbf{c}_{N})=e^{-\mu |\mathcal{A}|}\frac{\mu^N}{N!}\,d\mbf{c}_{N}\;,
\end{equation}
where $|\mathcal{A}|$ is the Lebesgue measure of $\mathcal{A}$ and $\mbf{c}_{N}=(c_1,\dots, c_N)$, $d\mbf{c}_{N}=dc_1\dots\,dc_N$.

One point particle is moving in the plane and bouncing among the obstacles, which keep their positions fixed. We refer to the classical Lorentz model for electron conduction in a random array of ions \cite{L05}. This model 
has a long history both in the physics and in the mathematical literature, see e.g.\,\cite{H,CD,S1,Sz}. In particular, it was considered by Gallavotti to give the first rigorous proof of the Boltzmann limit conjecture, as formulated by Grad \cite{Ga69,Ga70}; see \cite{S,LS,BBS,DP,BNPP,N,LT} for related results and subsequent developments. These works focus on the low-density (Grad) regime in which the intensity $\mu$ in Eq.\,\eqref{poisson} is rescaled as 
$$
\mu_{\varepsilon}=\varepsilon^{-1}\mu
$$
and $\e \to 0^+$.
We will denote below by $\PP_\ep, \EE_{\ep}$  the corresponding rescaled probability measure and expectation.

In the present paper the particle moves, between one collision and the next, under the action of a uniform, constant, magnetic field orthogonal to the plane. It is therefore subject to a force $F(v)=-Bv^{\perp}$
where $B>0$ is the magnitude of the magnetic field and $v^{\perp}=(v_2, -v_1)$, being $v = (v_1,v_2)$ the velocity of the particle.
To simplify notation, we set $\mu=1$ and assume that the particle has charge $q=-1$, mass $m=1$ as well as velocity of modulus $|v| = 1$. At contact with an obstacle, the particle is reflected elastically. Between two consecutive scatterers it moves counterclockwise with constant angular velocity $\Omega=-qB /{m}=B$ and performs an arc of circle of Larmor radius $R=|v|/B = 1/B$ (see Figure \ref{fig:1}). 
The cyclotron period is $T= 2 \pi /\Omega = 2\pi / B$.
For slight modifications of this model, a Markovian equation has been derived rigorously in \cite{MN}.
\begin{figure}[th]
\centering
\includegraphics [scale=0.45]{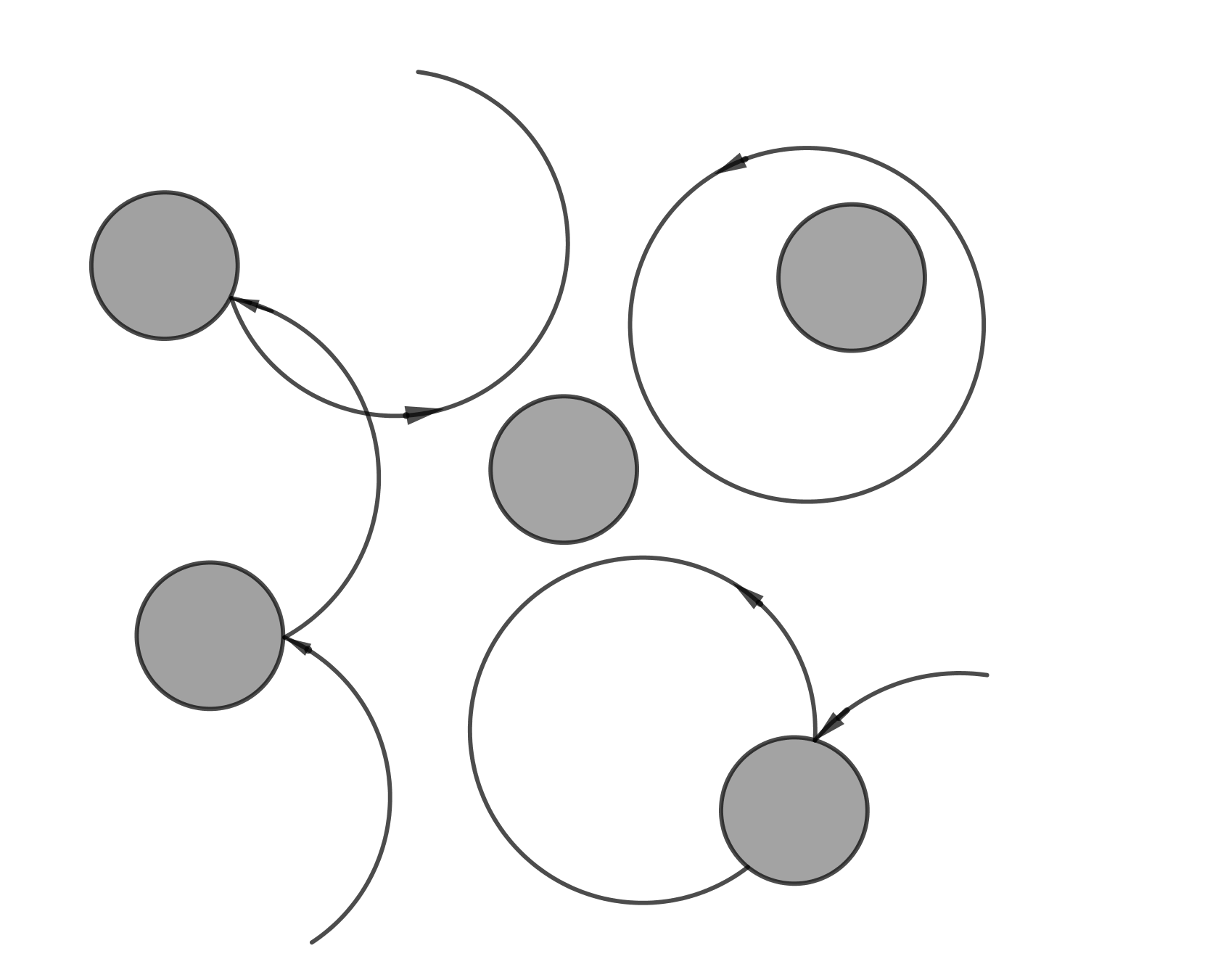}
\caption{Lorentz gas with magnetic field} \label{fig:1}  
\end{figure}

At finite densities, the transverse magnetic field produces a rich phenomenology due to the formation of traps consisting of Larmor orbits or clusters of scatterers (\cite{KS}). Moreover, the magnetic field affects the derivation of the Boltzmann equation in the limit of Grad. Indeed, Bobylev et al.\;have shown that both closed orbits and a certain class of recollisions are not negligible in this limit.

The following simple computation turns out to provide a good heuristic argument. Consider the probability 
of performing a complete cyclotron orbit starting from a given position and velocity $(x,v) \in \R^2\times S^1$ (where $S^1$ is the unit circle) without ever hitting any obstacle (circling particle).
One easily gets  
\begin{equation} \label{eq:circ}
\mathbb{P}_{\ep}(\{ \mathcal{C}\})=e^{-\mu_{\e} |\mathcal{A}_{\e}(R)|} = e^{- 4\pi R} = e^{-2T}
\end{equation}
where $\mathcal{C}$ is the event such that zero obstacles appear in the annulus $\mathcal{A}_{\e}(R)$ of radii $(R - \e, R + \e)$. Namely in the limit $\e\to 0$, there is a non vanishing probability for the particle to be a circling particle, simply due to the fact that the mean free path in the low-density regime is finite. 
Clearly this event is not described by the standard Boltzmann equation.
{More interestingly we will see that, with high probability, if there is one collision with an obstacle, there will follow infinitely many collisions with new obstacles.
But the same computation as in \eqref{eq:circ} shows that, after a collision,}
there is a finite probability of returning to a scatterer, via a cyclotron orbit, for additional encounter (recollision).
A non-Markovian structure is therefore not surprising, and the Boltzmann chaos breaks down as the absence of correlations prior to a collision fails.

\begin{figure}[th]
\centering
\includegraphics [scale=0.027]{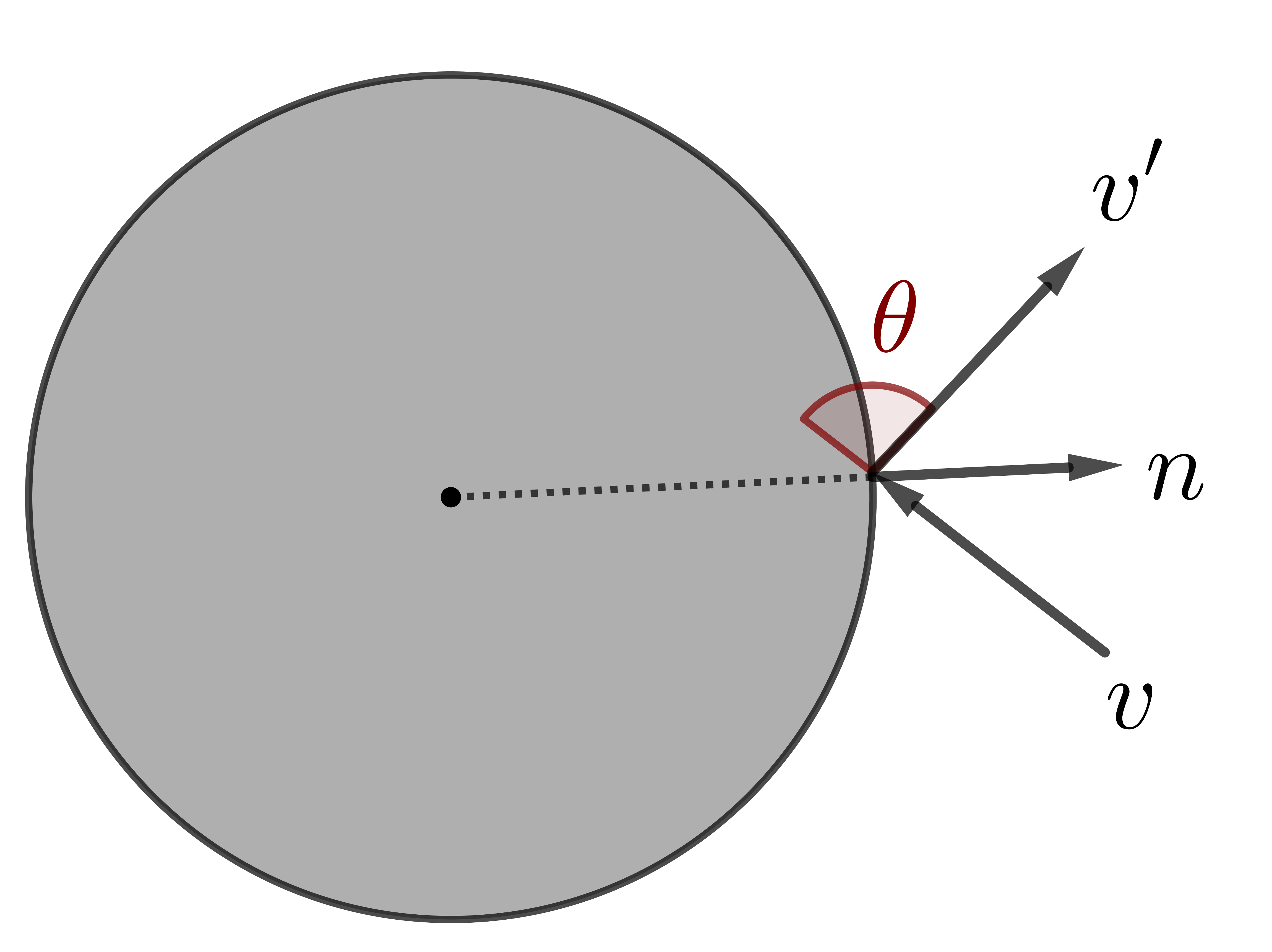} \caption{Scattering} \label{fig:scattering}  
\end{figure}

In order to take into account these effects, a linear kinetic evolution equation with memory, called {\it generalized Boltzmann equation}, has been derived and studied in \cite{BMHH, BMHH1,BHPH}. In the case of hard disks, it reads
\begin{equation}\label{eq:GBE}
\begin{split}
D_t f^{G}(t,x,v)=&
\sum_{k=0}^{[t/T]}e^{-2 k T}\int_{S^1}dn\,(v\cdot n)_+\,[ \sigma_n-1]\,f^{G}(t-kT,S_{n}^{(k)}(x,v))\;,\qquad t > 0\;.
\end{split}
\end{equation}
Here
$$D_t:=\left(\partial_{t}+v\cdot\nabla_{x}-(v\times B)\cdot\nabla_{v}\right)$$
is the generator of the free cyclotron motion with frequency $\Omega=B$ and $[t/T]$ is the number of cyclotron periods $T$ completed before time $t>0$, being $[x]$ the integer part of $x$. The symbol $(\cdot)_+$ denotes the positive part, and $n\in S^1$ is the scattering vector (see Figure \ref{fig:scattering}). In the gain term the operator $\sigma_n$ is defined by
$$\sigma_n\phi(x,v)=\phi(S_n(x,v)) $$
for arbitrary functions $\phi$, with the scattering map $S_n$ given by
\begin{equation} \label{eq:standardscatt}
S_n(x,v) := (x,v') \equiv  (x,v-2(v\cdot n) n)\;.
\end{equation}
The precollisional velocity $v'=v-2(v\cdot n)n$ becomes $v$ after the elastic collision with the hard disk.
Note that $v'\cdot n \leq 0$.
In the loss term, $v$ is precollisional with respect to~$-n$. 
Finally, $S_{n}^{(k)}(x,v)
:= \left(x,R_{k\theta}(v)\right)$, having denoted by $R_\alpha$ the $\alpha$-rotation 
\begin{equation}\label{eq:rotationR}
R_{\alpha} = \left (
\begin{array}{ll}
\cos(\alpha) &-\sin(\alpha)\\
\sin(\alpha) & \cos(\alpha)
\end{array}
\right)\;,
\end{equation}
and by  $\theta$ the scattering angle formed by $v-2(v\cdot n) n$ with respect to $v$.
In particular, $S_n^{(1)} = S_n$.

Only for $t < T$, the unknown $f^G$ should be interpreted as the probability density $f=f(t,x,v)$ of finding a test particle at time $t$ in position $x$ with velocity $v$. 
In fact, it is natural to split the total distribution function $f$ into two parts, $f = f^C+ f^W$, corresponding to {\it circling} particles and {\it wandering} particles respectively. 
The particle is trapped in a free cyclotron orbit of period $T$ with probability $e^{-2 T}$ (circling particle). With probability $1-e^{-2 T}$,
it collides with some obstacle before time $T$, and in that case it will collide with arbitrarily many different obstacles as predicted by \eqref{eq:GBE} (wandering particle). We therefore have
\begin{equation} 
f^{G}(t,x,v)=\left\{\begin{array}{ll}
f(t,x,v) = f^C(t,x,v) + f^W(t,x,v)\quad\quad\quad\quad\ \text{if}\;0<t<T&\\
f^W(t,x,v)\quad\quad\quad\quad\quad\quad\quad\quad\quad\quad\quad\quad\quad\quad\quad \text{if}\;t > T&
\end{array}\right.\;.\nonumber
\end{equation} 
While $\int_{\R^2\times S^1} f(t) = 1$ for all times, one has that 
\begin{equation} \label{eq:deffG}
\int_{\R^2\times S^1} f^G(t) = \left\{\begin{array}{ll}
1\quad\quad\quad\quad\quad\quad\quad\ \ \ \ \text{if}\;0<t<T&\\
1-e^{-2 T}\quad\quad\quad\quad\quad \text{if}\;t > T&
\end{array}\right.\;.
\end{equation} 
This renormalization condition has to be used in order to solve Eq.\,\eqref{eq:GBE} properly, as we will do in Section \ref{sec:GBE} below.

{The generalized Boltzmann equation is the forward equation of a process in which recollisions with a given scatterer (i.e.\,returns to the same collision point) happen with probability $1$; but only if no other scatterer has been hit in the meantime.
The admissible recollisions in the limiting process} are therefore chains of subsequent {\it self-recollisions}, i.e.\,subsequent collisions with the same obstacle, in which the scattering angle keeps always the same value (see Def.\,\ref{def:recollision} and Fig.\,\ref{fig:IntRec}   below). The sum over $k$ in \eqref{eq:GBE} takes into account the number of such possible recollisions, each one weighted by a factor $e^{-2 T}$ {(cf.\,\eqref{eq:circ}).}

Despite the fact that memory terms appear in the equation, the low-density limit significantly simplifies the phenomenology, as explained in \cite{BMHH}. 
The test particle is either a circling particle or a wandering particle, and it cannot be trapped in clusters of a finite number of scatterers.  

In this paper we address the question whether the generalized Boltzmann Eq.\,\eqref{eq:GBE} can be derived rigorously. Our aim is to provide an affirmative answer. We shall actually prove convergence of the particle process to the linear  process governed by Eq.\,\eqref{eq:GBE}.

We denote by $\gamma^{t}_{\mbf{c}_{N},\ep}(x,v)$, $t \in \R$ the Hamiltonian flow, solution to the Newton equations for a particle with initial configuration $(x,v)$, in a given sample $\mbf{c}_N$ of obstacles of radius $\e$ (cf.\,the formal definition \eqref{eq:ode-flow}-\eqref{eq:ode-flow'} below). For a given initial datum $f_0=f_0(x,v)$, the particle distribution function at time $t > 0$ is
\begin{equation}\label{eq:expect-density}
f^{\ep}(t,x,v):=\EE_{\ep}[f_0(\gamma^{-t}_{\mbf{c}_{N},\ep}(x,v))\, \mathbbm{1}_{\{\min_i |x-c_i| >\e\}}]\;.
\end{equation}

Furthermore for $t>0$, we denote the path space of the particle in $[0,t]$ by 
\begin{equation} \label{eq:pathsp}
\Pi_t = \bigcup_{m \geq 0} \Pi_{t,m}\;,
\end{equation} 
where $\Pi_{t,0}$ is the circling-path space (see Eq.\,\eqref{eq:circ}) and, for $m \geq 1$, the $m-$obstacle path space is defined by
\begin{equation} \label{eq:pathspm}
\Pi_{t,m} := \left\{ \left(t_1,n_1,\cdots,t_m,n_m\right)\ \big|\ 0 \leq t_1 \leq \cdots \leq t_m \leq t\;,\ n_i \in S^1\right\}\;.\end{equation}
The collection $\left(t_1,n_1,\cdots,t_m,n_m\right)$ provides the ingredients to describe the process, namely: the impact times 
 with new obstacles encountered by a particle starting from $(x_0,v_0)$ at time zero, and the scattering vectors at the moment of the encounter (cf.\,Definition \ref{def:times-vectors} below).

The path measure on $\Pi_t$ induced by the probability $\PP_\ep$, conditioned to the initial configuration $(x_0,v_0) \in \R^2\times S^1$ ($|c_i-x_0| \geq \e$ $\forall i$), will be indicated by $P_{\e,t}^{(x_0,v_0)}$. To fulfil \eqref{eq:deffG}, we cutoff circling particles after the first Larmor period, i.e.\,we set $P_{\e,t}^{(x_0,v_0)}\left(\Pi_{t,0}\right) = 0$ for $t > T$. 
We shall fix $(x_0,v_0)$ from now on, and refer to the random trajectory $\left(\zeta^{\ep}(s)\right)_{s\in[0,t]}$ starting from $(x_0,v_0)$ as the {\it Lorentz process} (cf.\,Definition \ref{def:flow} below).
 
Since $\zeta^{\ep}$ has jumps in velocity, we work in the path space $D\left([0,t], \R^2\times S^1\right)$ of piecewise continuous functions equipped with the Skorokhod topology (see \cite{Bi70}).
Let $\left(\zeta(s)\right)_{s\in[0,t]}$ be the  generalized Boltzmann process in $D\left([0,t], \R^2\times S^1\right)$ with forward equation \eqref{eq:GBE}-\eqref{eq:deffG}{, starting from $(x_0,v_0)$ (cf.\,Definition \ref{def:BP} below). 
 Informally, this Boltzmann process is described as follows. Pick a point in $\Pi_t$ (say, $t>T$). If $m=0$, the particle is trapped on a Larmor orbit. Otherwise if $m \geq 1$, then necessarily $t_1 \leq T$ and at time $t_1$ a Boltzmann-type collision  occurs with scattering vector $n_1$ and scattering angle $\theta_1$. From that time on, new Boltzmann-type collisions will occur at times $t_2, t_3\cdots$ with scattering determined by $n_2, n_3\cdots$. In between these collisions, the process will keep moving on Larmor orbits, coming back to the $i$th collision point $[|t_{i+1}-t_i|/T]$ times. Each of such revisits is a self-recollision, leading to a scattering of the same angle $\theta_i$, and making the particle jump to a correspondingly rotated orbit (see also Figure \ref{fig:1} below).
}

{
Our main result states the convergence of the Lorentz process to the generalized Boltzmann process in the annealed setting, i.e.\,averaging over the random (Poisson) placement of the scatterers.
}

\begin{theorem}\label{thm:main}
 For all $t>0$, we have that 
   \begin{equation} \label{eq:thm}
\lim_{\e \to 0}\left\{s \to\zeta^{\ep}(s)\right\}_{s\in[0,t]}= \left\{s \to\zeta(s)\right\}_{s\in[0,t]}
  \end{equation}
in the sense of weak convergence of path measures on $D\left([0,t], \R^2\times S^1\right)$.

For any probability density $f_0\in C(\R^2\times S^1)$, the particle density distribution $f^\e(t)$ defined in \eqref{eq:expect-density} satisfies
\begin{equation} \label{thm:cor}
\lim_{\e\to 0}\| f^\e(t)-f(t)\|_{L^1(\R^2\times S^1)}=0\;,
\end{equation}
where $f$ is the unique mild solution to \eqref{eq:GBE}-\eqref{eq:deffG} in $C(\R^2\times S^1)$ with initial datum $f_0$.  
\end{theorem} 

Our purpose is to present a simple argument. To this end: \\
(i) We have considered only scatterer configurations which interact with the particle as hard disks. The method works as well for more general short range potentials, however additional difficulties arise from a strictly positive (for $\e >0$) scattering time and from singularities in the differential cross-section. Notice that, for smooth potentials, the differential cross section depends also on $B$ as soon as $\e>0$.\\
(ii) Theorem \ref{thm:main} is stated without any explicit rate of convergence. We leave an analysis of the corrections to future work.\\
(iii) A more general class of scatterer distributions, not necessarily Poisson, could be discussed as we do in this paper, with minor changes. We refer to \cite{S1} for the required assumptions.

The proof of Theorem \ref{thm:main} is based on an extension of the arguments in \cite{Ga69,Ga70}. The basic ingredient is a suitable parametrization in terms of impact times and impact vectors, coupling the Lorentz process with the limiting process. The main difference here is that we need to deal with memory terms which do not vanish in the limit and that are characterized geometrically in terms of self-recolliding trajectories. We therefore need additional care in the coupling process and in its link with the formula \eqref{eq:GBE}.

We conclude the introduction by recalling that, in the case of periodic (deterministic) configurations of scatterers, the validity of the Boltzmann equation in the limit of Grad is known to fail \cite{CG1,CG2,M,MS}. The model considered in this paper is an instance of the fact that the background randomness may be not enough to ensure a Markov property, and that an external force field can strongly affect the asymptotic behaviour. Other examples of this feature have been studied in \cite{DR1,DR2}.

Finally, as a word of warning concerning the ``non-Markovian'' effects we are referring to (a terminology that we inherit from \cite{BMHH, BMHH1, BHPH}), we wish to stress that an underlying Markovian process still exists, given by the sequence of path segments between different encountered obstacles. Our method exploits this fact by working on the path space $\Pi_t$ parametrizing ``fresh'' collisions. The proposed Lorentz model can be therefore regarded as a simple way to add memory (of unbounded order) on top of this Markovian structure.
As further noticed in \cite{BMHH1}, it is possible to recover an analogous result by considering the classical random Lorentz Gas (without magnetic field) on the $S^2$ sphere.

The paper is organized as follows: Section \ref{sec:dynamics} contains a preliminary description of the mechanical process; the generalized Boltzmann process governed by Eq.\,\eqref{eq:GBE} is studied in Section \ref{sec:GBE}; Section \ref{sec:proof} is devoted to the proof of Theorem \ref{thm:main}.

\section{The Lorentz process}\label{sec:dynamics} 
 
The particle motion takes place along cyclotron (circular) orbits
and collisions make the particle pass from one orbit to the next. That is, as a result of
the particle colliding, the center of the cyclotron orbit jumps to a different
position. 

\begin{figure}[th]
\centering
\includegraphics [scale=1.17]{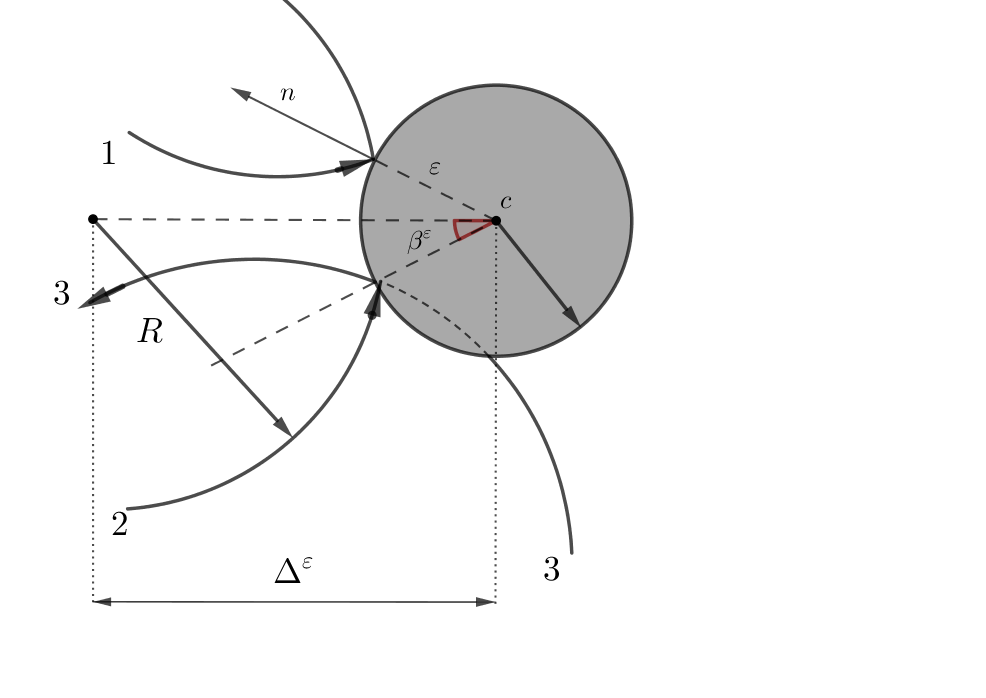} 
\caption{A trajectory with one collision and two self-recollisions} 
\label{fig:2}  
\end{figure}

Consider for instance the case of three subsequent collisions with the same obstacle: see Figure \ref{fig:2}. 
Let the initial cyclotron orbit be the one labelled by $1$. In the first encounter with the hard disk,
the particle is scattered over an angle $\theta$ and switches to the cyclotron orbit
labelled by $2$. From the symmetry of the event it is clear that the cyclotron orbit $2$
has its center at the same distance $\Delta^\e$ from the center of the scatterer as the orbit $1$. Consequently, the orbit $2$ intersects the scatterer in precisely the
same way as the orbit $1$, only rotated by an angle $2\beta^\e$ along the circumference
of the disk (see Definition \ref{def:self-rec} below). 

 As observed in \cite{BMHH1}, with probability 1, during a long sequence of self-recollisions, the orbits will densely fill a ring shaped area around the scatterer with
outer radius $\Delta^\e+R$. Moreover, in the Boltzmann-Grad limit an important simplification
arises: on the length scale set by the size
of the scatterer, the cyclotron orbits degenerate into a straight line, and the accumulated scattering angle after $k$ self-recollisions equals exactly $(k+1)\theta$.

We turn now to a more formal description.

\begin{definition}[Lorentz process]\label{def:flow}
Given $\e >0$, $(x_0,v_0) \in \R^2 \times S^1$ and a random configuration of scatterers $\mbf{c}_N$ such that $\min_i |c_i - x_0| \geq \e$, the Lorentz process $t \to \zeta^\e(t)$ starting from $(x_0,v_0)$ is given by 
\begin{equation} \label{eq:zet}
\zeta^\e(t) \equiv \left(\xi^\e(t),\eta^\e(t)\right)=\gamma_{\mbf{c}_N,\e}^{t}(x_0,v_0)\;,\qquad t \in \R\;,
\end{equation} 
where $\gamma_{\mbf{c}_N,\e}^{t}$ is defined almost surely on
\begin{equation*} \label{eq:LecN}
\Lambda^\e_{\mbf{c}_N}:=\{(x,v)\in\R^2\times S^1\ |\ \min_i |x-c_i| >\e \}
\end{equation*} 
by
\begin{equation}\label{eq:ode-flow}
\left\{
\begin{array}{ll}
\dot{\xi}^\e(t)=\eta^\e(t)  \\
\dot{\eta}^\e(t)=-B \left(\eta^\e(t)\right)^{\perp}  \\
\left(\xi^\e(0),\eta^\e(0)\right)=(x_0,v_0) \\
\end{array}
\right.\;,
\end{equation}
with the boundary conditions on $\partial\Lambda^\e_{\mbf{c}_N}$:
\begin{equation} \label{eq:ode-flow'}
\eta^\e \longrightarrow \left(\eta^\e\right)' = \eta^\e - 2 \left(\eta^\e\cdot \left(\frac{\xi^\e - c_i}{\e}\right)\right) 
\frac{\xi^\e - c_i}{\e}\qquad \mbox{if$\quad |\xi^\e - c_i| = \e$}\;.
\end{equation} 
The case $N=0$ corresponds to the collisionless flow
$\left(\xi^\e(t),\eta^\e(t)\right)=\gamma_{\emptyset,\e}^{t}(x_0,v_0) \equiv \gamma_{0}^{t}(x_0,v_0)$
defined by \eqref{eq:ode-flow} over $\R^2 \times S^1$ with no boundary conditions.
\end{definition}
Notice that, with probability zero, the particle can hit two obstacles simultaneously: in which case the process is ill-defined. Otherwise, \eqref{eq:ode-flow'} is an involution from $\partial\Lambda^{\e-}_{c_i}$ to $\partial\Lambda^{\e+}_{c_i}$ where
$$
\partial\Lambda^{\e \pm}_{c_i} := \{(x,v)\in\partial\Lambda^{\e}_{c_i}\ |\ \pm v \cdot \left(x-c_i\right) \geq 0\}\;.
$$
The so defined process moves on a (possibly finite, possibly infinite) sequence of Larmor orbits of radius $R = 1/B$. 

Almost surely, each trajectory in $[0,t]$ is in one-to-one correspondence with a finite collection of impact times and vectors, which we introduce next. We first single out the obstacles responsible for collisions.
\begin{definition}[Internal and external obstacle]\label{def:int-ext}
An obstacle is internal if its center $c_i$ is such that 
\[
\inf_{t \geq 0}|\xi^\e(t)-c_i|=\ep,
\]
while it is external if
\begin{equation*}
\inf_{t \geq 0}|\xi^\e(t)-c_i|>\ep\;.
\end{equation*}
\end{definition}
\begin{definition}[Impact time and impact vector]\label{def:times-vectors}
The impact time of an internal obstacle with center $c_i$ is defined as
\[
t_i := \sup\left\{\tau>0\ \Big|\ \inf_{0\leq t\leq \tau}|\xi^\e(t)-c_i|>\e\right\}\;.
\]
The impact vector of an internal obstacle with center $c_i$ is defined as
\[
n_i := \frac{\xi^\e(t_i) - c_i}{\e} \in S^1\;.
\]
We say that internal obstacles are ordered if $0<  t_1 < t_2  \cdots $.\\
\end{definition}

The phenomenon preventing molecular chaos, namely the statistical  independence in the limit, are the so-called recollisions, defined as follows.
\begin{definition}[Recollision] \label{def:recollision}
Collisions occurring at times different from impact times are called recollisions.
A recollision with a scatterer labelled by $i$ at time $\tau$ is a self-recollision if the last scattered obstacle before $\tau$ is $i$ itself.
\end{definition}
\begin{figure}[th]
\centering
\includegraphics [scale=0.08]{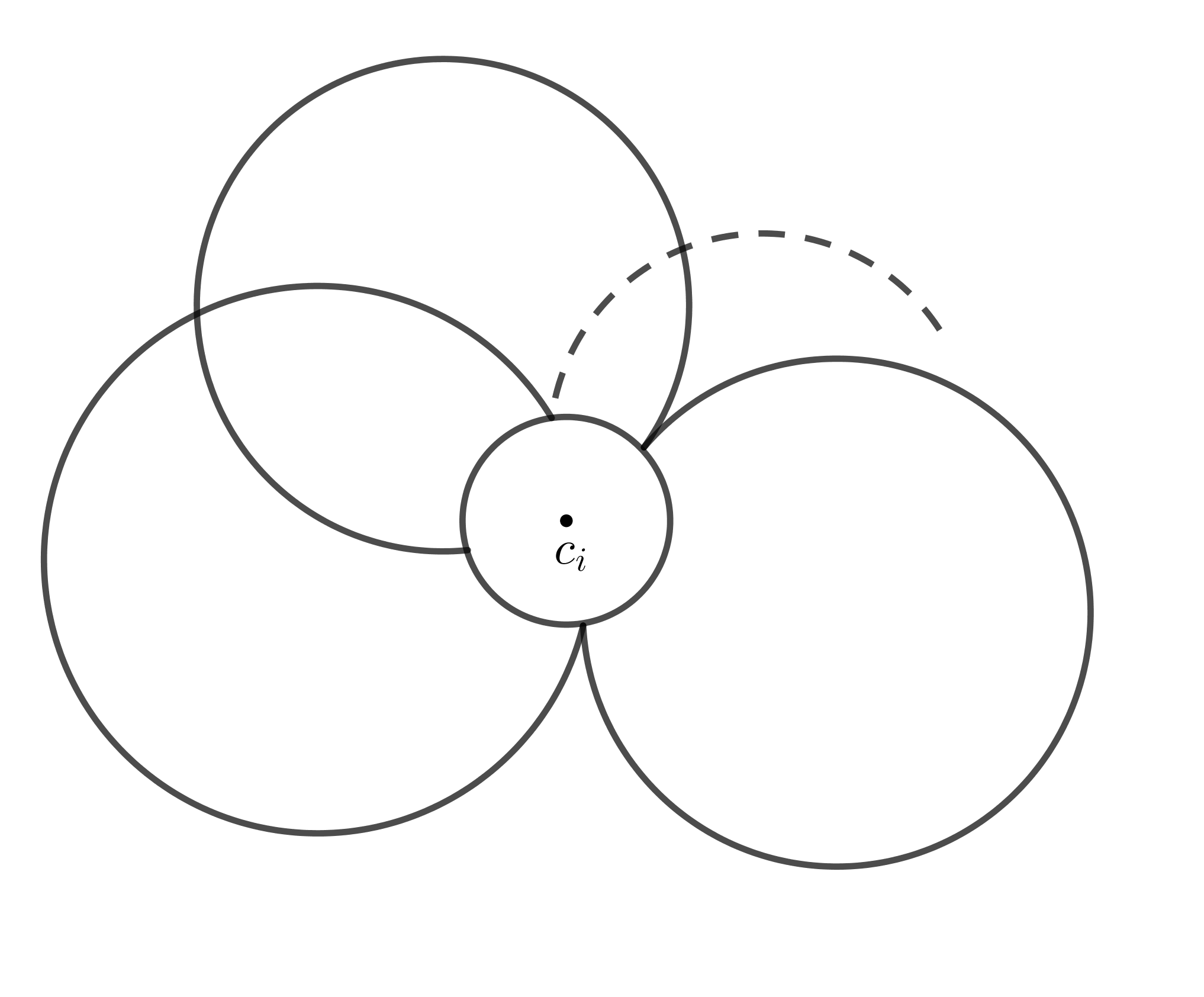} 
\centering\caption{Self-recollision: if no new obstacle is encountered in a Larmor period $T$, 
then the trajectory recollides with the last scattered obstacle.} \label{fig:IntRec}  
\end{figure}

\begin{figure}[th]
\centering
\includegraphics [scale=0.17]{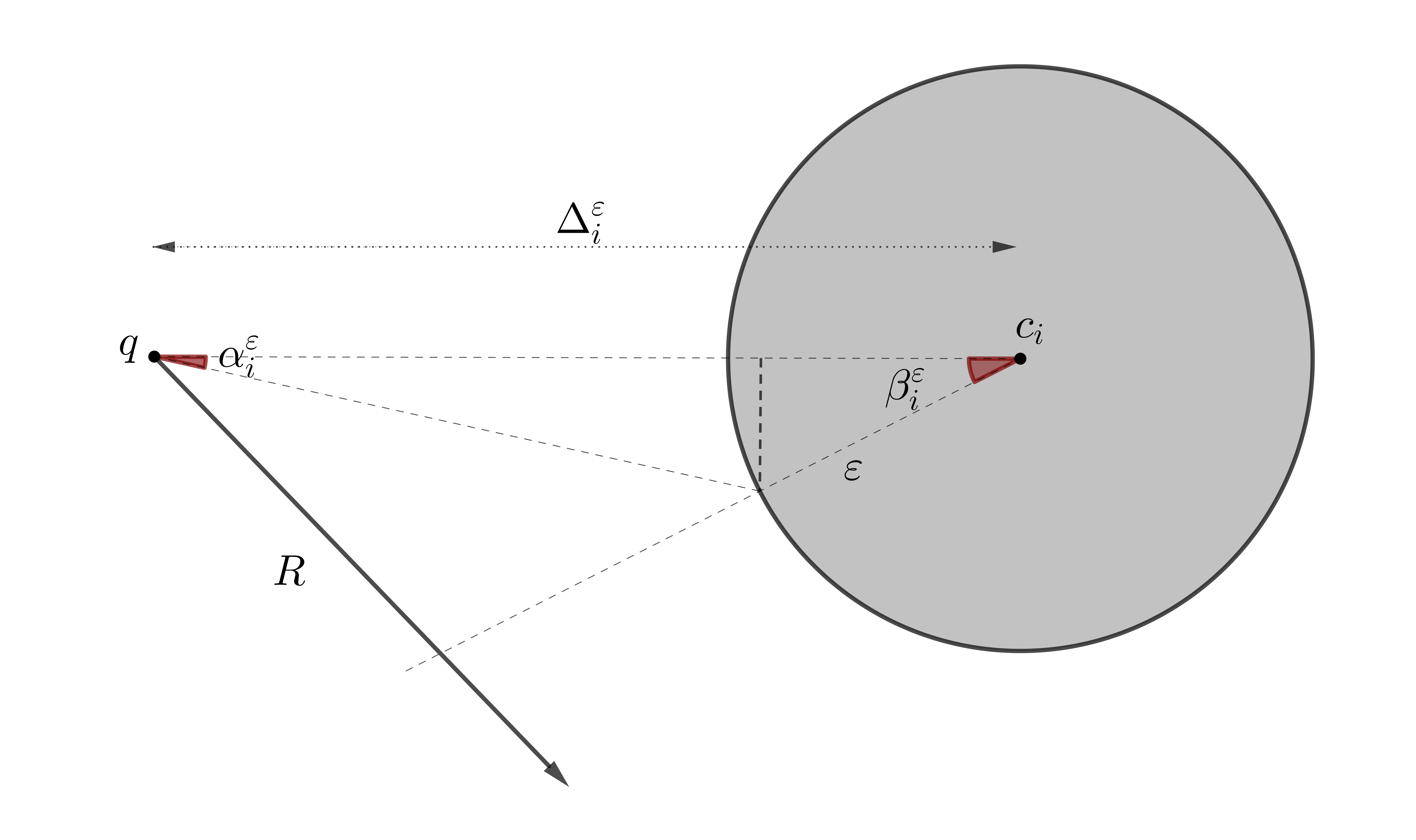} 
\caption{Geometry of a self-recollision.} 
\label{fig:2bis}  
\end{figure}

As mentioned above, we will prove that self-recollisions are not negligible in the limit $\e \to 0$. It is therefore important to track this class of recollisions precisely. The following map describes the jumps in position and velocity, leading to a self-recollision.
\begin{definition}[Self-recollision map]\label{def:self-rec}
Let $c_i \in \R^2$ be the center of an internal scatterer. The self-recollision map is the $\e$-dependent map 
$S^\e : \partial\Lambda^{\e -}_{c_i} \to \partial\Lambda^{\e -}_{c_i}$ defined by  (cf.\,\eqref{eq:rotationR})
%
\begin{equation*}\label{eq:eps-scatter+shift}
S^\e(c_i + \e\, n_i,v):=\left(c_i +\e\, R_{2 \beta^\e_i}\left(n_i\right), \ R_{\theta^\e_i}(v)\right)\;,
\qquad n_i \in S^1\;,\ v\cdot n_i \leq 0
\end{equation*}
where: 

\noindent (i) $\beta_i^{\e}>0$ is given by
%
\begin{equation*}\label{eq:beta}
\cos \beta_i^\e=\frac{(\Delta_i^{\e})^2-R^2+\e^2}{2\Delta_i^\e \e}\;,
\end{equation*}
with $\Delta_i^{\e} := |q - c_i| \in [R-\e,R+\e]$ where $q$ is the center of the Larmor orbit spanned by a particle in $(c_i + \e n_i,v)$;

\noindent (ii)
$\theta_i^\e \in (-\pi,\pi]$ is  given by 
$$\theta_i^\e = \theta_i - 2 \alpha_i^\e$$
with $\theta_i$ the scattering angle of the collision in $(c_i + \e n_i,v)$ (formed by $v-2(v\cdot n_i) n_i$ with respect to $v$) and 
$$\sin\alpha_i^\e=\frac{\e}{R}\sin\beta_i^{\e}\;.$$
\end{definition}
See Figure  \ref{fig:2} and Figure \ref{fig:2bis} for the main ingredients appearing in the above definition.
Observe that 
\begin{equation*}
R_{\theta^\e_i}(v) \cdot R_{2 \beta^\e_i}\left(n_i\right) = v \cdot n_i \leq 0\;,
\end{equation*}
hence the map $S^\e$ leads to a further collision with the obstacle, with scattering angle $\theta_i$.

Since $\theta_i^\e \to  \theta_i$ as $\e$ tends to $0$ and the scatterer degenerates into a point, the self-recollision map converges to the standard scattering map with unchanged position (cf.\,Eq.\,\eqref{eq:standardscatt}): 
$$\lim_{\e \to 0 }S^\e(c_i + \e\, n_i,v) = S_{n_i}(c_i,v)\;.$$
Actually since $\theta_i^\e = \theta_i + O(\e)$, one has that
\begin{equation}\label{eq:err}
S^\e(c_i + \e\, n_i,v) = S_{n_i}(c_i,v) + O(\e)\;.
\end{equation}
Further details on this map and a discussion in the (more complicated) case of smooth interactions can be found in \cite{MN}.

In a pictorial language, we can say that a self-recollision process around an obstacle has a trajectory spanning a daisy shape around it. Each self-recollision corresponds to one daisy petal.
\begin{definition}[Daisy]\label{def:daisy}
Let $\zeta^\e(t)$ be a Lorentz process with ordered internal obstacles centered in $c_1,c_2,\cdots$.
A daisy around the internal obstacle labelled by $i$ is the region
$$\mathcal{D}^\e_{i}:= \Big\{ x\in\R^2\ \Big|\ \inf_{s \in (0,t_{i+1}-t_i)}|x-\gamma_{c_i,\e}^{s}\left(\zeta^\e(t_i)\right)| < \e \;,\ \min_{j=i,i+1} |x-c_j| >\e \Big\}\;.$$
The stem of the first daisy is
$$
\mathcal{D}^\e_{0} := \Big\{ x\in\R^2\ \Big|\ \inf_{s \in (0,t_1)}|x-\gamma_{0,\e}^{s}\left(\left(x_0,v_0\right)\right)| < \e \;,\ |x-c_1| >\e \Big\}\;.
$$
\end{definition}
Note that in the definition of $\mathcal{D}^\e_{i}$ we used the $1$-obstacle flow $\gamma_{c_i,\e}$ instead of \eqref{eq:zet}, to describe the trajectory in $(t_i,t_{i+1})$. This means that we are disregarding recollisions with different obstacles which might be present in the Lorentz process and deviate the trajectory from the daisy. We will however prove in Section \ref{sec:proof} that these recollisions have vanishing probability as $\e \to 0$.

\section{The generalized Boltzmann process}\label{sec:GBE}

Our purpose in this section is to solve the generalized Boltzmann equation \eqref{eq:GBE}-\eqref{eq:deffG}, introduced in \cite{BMHH}, and find an explicit formula for the path measure $P_t^{(x_0,v_0)}$ of the corresponding process. In particular, we search for a solution representation which can be coupled effectively with the Hamiltonian flow. This requires a discussion on the memory terms.

The main result of this section is  Proposition \ref{prop:Boltzmann} below. Before stating the result, we need new notations to describe the limiting trajectories. These are collected in the next definition.

We denote here by 
$\left(\xi_0(s),\eta_0(s)\right)_{s \in \R}=\left(\gamma_{0}^{-t_0+s}(x,v)\right)_{s \in \R}$ the collisionless flow with starting configuration $(x,v) \in \R^2 \times S^1$ at time $t_0 \in \R$,
defined on $\R^2 \times S^1$ and for any $t_0 \in \R$ as the solution of the following system of ODEs 
\begin{equation}\label{eq:cpf}
\left\{
\begin{array}{ll}
\dot{\xi}_0(s)=\eta_0(s)  \\
\dot{\eta}_0(s)=-B \left(\eta_0(s)\right)^{\perp}  \\
\left(\xi_0(t_0),\eta_0(t_0)\right)=(x,v) \\
\end{array}
\right.\;.
\end{equation}
In order to solve the Boltzmann equation we only need a backward flow, however we shall define both a backward and a forward flow, for future convenience. 
\begin{figure}[th]
\centering
\includegraphics [scale=0.54]{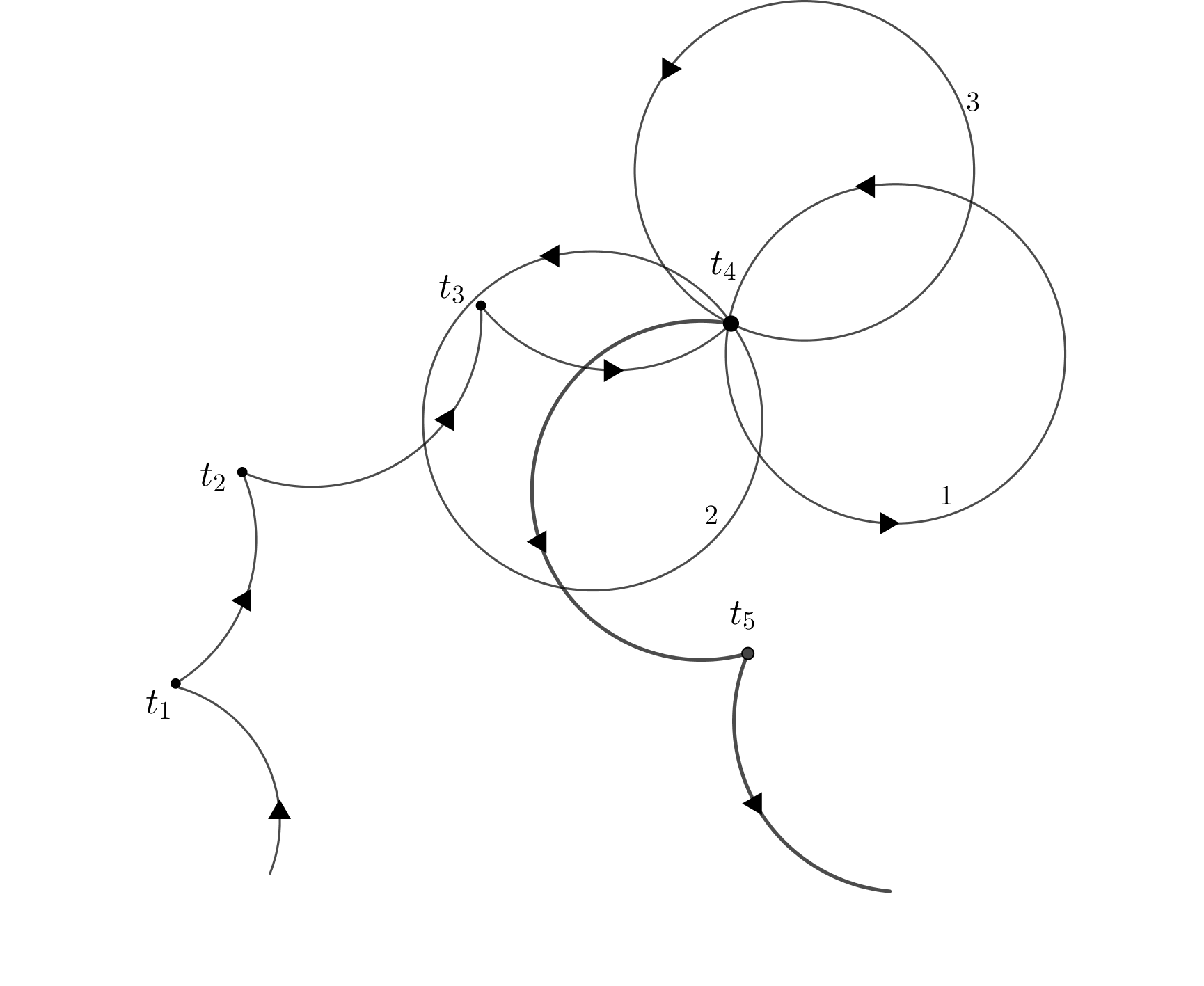}
\caption{{ Generalized (forward) Boltzmann flow; $m=5$, $k_4=3$, $k_i=0$ for $i \neq 4$. The self-recollision orbits are ordered as in Figure \ref{fig:2}.}} \label{fig:1}  
\end{figure}
\begin{definition}[Boltzmann flow] \label{def:BFlow}
A generalized Boltzmann flow $\zeta_m(s)$  in $D\left(\R, \R^2\times S^1\right)$ with starting configuration $(x,v) \in \R^2 \times S^1$ at time $t_0 \in \R$, and $m \in \N$ obstacles (or collisions, modulo possible self-recollisions), is 
\begin{equation} \label{eq:GBF}
\zeta_m(s) \equiv \left(\xi_m(s),\eta_m(s)\right)=\gamma_m^{-t_0+s}(x,v)\;,	\qquad s \in \R\;,
\end{equation}
where $\gamma_m^{-t_0+s}$ is constructed as follows.

For all $m$, $\gamma_m^{0}(x,v) = (x,v)$.
The flow $\gamma^{-t_0+s}_0$ is already defined by \eqref{eq:cpf}. 
The flows $\gamma_m^{-t_0+s}(x,v) $, $m \geq 1$, are
defined iteratively.
\begin{description}
\item[1] Choose $t_1$ such that $0 \leq t_1-t_0 \leq T $ if $s>t_0$, and $0 \leq t_0-t_1 \leq T $ if $s<t_0$. Apply $\gamma_0$ to the configuration $(x,v)$,  up to the time $t_1$ (forward Larmor orbit if $s>t_0$, backward Larmor orbit if $s<t_0$): this defines  $\zeta_m(s) = \left(\xi_m(s),\eta_m(s)\right)$ for $s$ from $t_0$ up to $t_1$. 
\item[2] 
Start with the configuration obtained in the previous step ($i \geq 1$). This configuration is $\zeta_m(t_{i}^-)$ if $s > t_0$, or  $\zeta_m(t_{i}^+)$ if $s < t_0$. Choose $n_i \in S^1$ such that $n_i\cdot\eta_m(t_{i}^-) \leq 0$ if $s>t_0$, or $n_i\cdot\eta_m(t_{i}^+) \geq 0$ if $s<t_0$. Apply the scattering map $S_{n_i}$, which rotates the velocity of an angle $\theta_i$: this defines a configuration $\zeta_m(t_{i}^+)$ if $s > t_0$, or  $\zeta_m(t_{i}^-)$ if $s < t_0$, which is the starting point for the following step.
\item[3] Choose $t_{i+1}$ such that $ (t_{i+1}-t_i) (s-t_0) \geq 0$. Set 
\begin{equation} \label{eq:defki}
k_i := [|t_{i+1}-t_i|/T]\;.
\end{equation} 
Apply $\gamma_0$ until the first Larmor orbit is completed (at time $t_i + T$ or $t_i - T$, for $s>t_0$ or $s<t_0$ respectively), then rotate the velocity of an angle $\theta_i$: repeat this step for $k_i$ times up to time $t_i \pm k_iT$. This describes $k_i$ Larmor orbits rotated of the same scattering angle around the point $\xi_m(t_i)$ (self-recollisions of the Boltzmann flow).
\item[4] Apply $\gamma_0$ from the time $t_i \pm k_iT$ up to the time $t_{i+1}$. Return to item {\bf 2}.
\end{description}
If the condition on the sign of $n_i\cdot\eta_m(t_{i}^-)$ is not respected, we extend the above definition by setting $\theta_i = 0$.
\end{definition}
Note that the generalized Boltzmann flow with $m$ obstacles restricted to $[0,t]$ is everywhere defined on $\Pi_{t,m}$, for all $t$ (cf.\,\eqref{eq:pathspm}).

Let $\chi(A)$ denote the characteristic function of the set $A$, and $a\wedge b = \min(a,b)$.
\begin{proposition}\label{prop:Boltzmann}
Let $f^C$ be the density of circling particles
   $$
 f^C(t,x,v) := e^{-2 t \wedge T}\,f_0(\gamma_0^{-t \wedge T}(x,v))
 $$
 where we use Eq.\,\eqref{eq:cpf} in the case $t_0 = t$. Furthermore define
  \begin{equation}
 \label{eq:Bmss}
\begin{split}
f^G (t,x,v) &:= \chi(\{  t<T\}) \, f^C(t,x,v) \\&
+ e^{-2 t}\int_{(t-T) \wedge 0}^{t}
dt_1 \int_{S^1} dn_1 \sum_{m \geq 1}  \int_0^{t_1} dt_2 \cdots \int_0^{t_{m-1}}dt_m \int_{\left(S^{1}\right)^{m-1}} dn_2\cdots dn_m\\&
\qquad
\cdot\left[ \prod_{i=1}^m  \left(n_i \cdot \eta_m(t_i^+)\right)_+ \right] f_0(\gamma_m^{-t}(x,v))\;,
\end{split}
 \end{equation}
where we set $t_{m+1}=0$ in the definition of $\gamma_m$ (Def.\,\ref{def:BFlow}).
Then, $f^G$ is the unique mild solution to \eqref{eq:GBE}-\eqref{eq:deffG} in $C(\R^2\times S^1)$ with initial datum $f_0$. 
\end{proposition}
Eq.\,\eqref{eq:Bmss} provides an explicit solution. 
 As required, the solution coincides with the solution of the true Boltzmann equation with magnetic field only for $t<T$: it is in this case identical to the formula obtained in \cite{Ga69}, except for the fact that straight line trajectories are replaced by arcs of Larmor orbits. For $t>T$, the non-Markovian character arises, and self-recollisions are possible as described in item {\bf 3} in Definition \ref{def:BFlow}.
 
Note that, exactly as in \cite{Ga69}, Eq.\,\eqref{eq:Bmss} has a simple probabilistic interpretation as integral over paths. The right hand side of the first line describes the circling particles, while the second and third lines describe the wandering particles. 
Once the particle hits the first obstacle, it will be wandering forever: the subsequent obstacles can be encountered at arbitrary times. For this reason, there is no constraint (apart from the order) in the integrals over $t_2,\cdots,t_m$.
 For further analysis of the solution we refer to \cite{BMHH, BMHH1, BHPH}.

\bigskip
\noindent {\em Proof of Proposition \ref{prop:Boltzmann}}. We start by rewriting Eq.\,\eqref{eq:GBE} as follows:
 \begin{align} \label{eq:GBE_2} 
&\partial_t f^G (t,x,v) = \Big(-v\cdot \nabla_x + (v \times B)\cdot \nabla_v -2\Big) f^G (t,x,v)
\nonumber \\&
+  \int_{S^1} dn   \left(n \cdot v\right)_+ 
 f^G(t,S_{n}(x,v))\nonumber
\\& +  \sum_{k \geq 1} \chi\left(\{t-kT \geq 0\}\right) e^{-2 kT}  \int_{S^1} dn   \left(n \cdot v\right)_+ \nonumber 
\\
&\;\;\;\cdot
\left\{ f^G(t-kT,S^{(k+1)}_{n}(x,v)) - f^G(t-kT,S^{(k)}_{n}(x,v)) 
\right\}\;. \nonumber
 \end{align}   
Imposing \eqref{eq:deffG}, we write the Duhamel formula for $0<t<T$ (standard Boltzmann equation)
 and then for $t >T$ (Boltzmann equation with memory) with the new initial data $$f^G(T^+,x,v) := f^G(T^-,x,v) - 
 f^C(T,x,v)$$
  where the notation $T^+$, $T^-$ in the argument indicates that the limit $t \to T$ is taken from the future, past, respectively. 
For all positive $t \neq T$:
 \begin{equation*}
\begin{split}
&  f^G (t,x,v) = \chi(\{ t<T\}) e^{-2 t} f_0(\gamma_0^{-t}(x,v)) \\
& \;\;\;  + 
\int_{0}^{t} ds \int dn \left(n \cdot \eta_0(s)\right)_+
e^{-2 (t-s)} f^G(s,S_{n}\gamma_0^{-t+s}(x,v)) 
\\
&\;\;\; +  \sum_{k \geq 1} \int_{k T \wedge t}^t ds \,e^{-2 (t-(s-kT))}  \int dn \left(n \cdot \eta_0(s)\right)_+\\
&\;\;\;\;\;\;\cdot\left\{ f^G(s-kT,S^{(k+1)}_{n}\gamma_0^{-t+s}(x,v)) - f^G(s-kT,S^{(k)}_{n}\gamma_0^{-t+s}(x,v)) 
\right\} \;.
\end{split}
 \end{equation*}
 Here and afterwards we use the shortened notation $\int = \int_{S^1}$.
 
The latter equation contains negative terms but,
with the help of simple changes of variables, we can rewrite it as a sum of positive contributions:
 \begin{equation*}
\begin{split}
& f^G (t,x,v)=  \chi(\{ t<T\})\, e^{-2 t} \,f_0(\gamma_0^{-t}(x,v)) \\&\;\;\;+ 
\int_{0}^{t} ds \int dn  \left(n \cdot \eta_0(s)\right)_+
e^{-2 (t-s)} f^G(s,S_{n}\gamma_0^{-t+s}(x,v)) 
\\
& \;\;\;+  \sum_{k = 1}^{^{[t/T]}} \int_{0}^{t - k T} ds\, e^{-2 (t-s)}  \int dn \left(n \cdot \eta_0(s)\right)_+\\
&\;\;\;\;\;\;
\left\{ f^G(s,S^{(k+1)}_{n}\gamma_0^{-t+s}(x,v)) - f^G(s,S^{(k)}_{n}\gamma_0^{-t+s}(x,v)) 
\right\}\\
& =  \chi(\{ t<T\})\,e^{-2 t}\, f_0(\gamma_0^{-t}(x,v)) \\&\;\;\;
+\int_{(t-T) \wedge 0}^{t} ds \int dn \left(n \cdot \eta_0(s)\right)_+
e^{-2 (t-s)} f^G(s,S_{n}\gamma_0^{-t+s}(x,v)) 
\\ 
&\;\;\; +   \int_{0}^{(t -  T) \wedge 0} ds \,e^{-2 (t-s)}  \int dn \left(n \cdot \eta_0(s)\right)_+
f^G(s,S^{(2)}_{n}\gamma_0^{-t+s}(x,v)) \\
&\;\;\; + \sum_{k = 2}^{^{[t/T]}} \int_{0}^{t - k T} ds\, e^{-2 (t-s)}  \int dn \left(n \cdot \eta_0(s)\right)_+\\
&\;\;\;\;\;\;
\left\{ f^G(s,S^{(k+1)}_{n}\gamma_0^{-t+s}(x,v)) - f^G(s,S^{(k)}_{n}\gamma_0^{-t+s}(x,v)) 
\right\}\\
& =  \chi(\{ t<T\})\,e^{-2 t}\, f_0(\gamma_0^{-t}(x,v))  \\&\;\;\;
+\int_{(t-T )\wedge 0}^{t} ds \int dn \left(n \cdot \eta_0(s)\right)_+
e^{-2 (t-s)} f^G(s,S_{n}\gamma_0^{-t+s}(x,v)) 
\\ 
&\;\;\; +   \int_{(t -  2T )\wedge 0}^{(t -  T) \wedge 0} ds\, e^{-2 (t-s)}  \int dn \left(n \cdot \eta_0(s)\right)_+
f^G(s,S^{( 2)}_{n}\gamma_0^{-t+s}(x,v)) +  \int_{(t -  3T )\wedge 0}^{(t -  2T) \wedge 0}\cdots\;.
\end{split}
 \end{equation*}
 Equivalently, using that $\gamma_0^{-t+s} = \gamma_0^{-t+s-kT}$ for all $k \in \N$ and that $S^{(k_1+1)}_{n_1}\gamma_0^{-t+t_1}(x,v) = \gamma_1^{-t+t_1-k_1T}(x,v)$ (cf.\,Definition \ref{def:BFlow}),
 \begin{equation} \label{eq:Duhfm}
\begin{split}
& f^G (t,x,v)=\chi(\{ t<T\}) \,e^{-2 t} \, f_0(\gamma_0^{-t}(x,v)) \\
&+ 
\int_{(t-T) \wedge 0}^{t} dt_1 \int dn_1  \left(n_1 \cdot \eta_0(t_1^+)\right)_+
e^{-2(t-t_1)} \sum_{k_1 = 0}^{[t/T]}  e^{-2 k_1 T} f^G(t_1-k_1T, \gamma_1^{-t+t_1-k_1T}(x,v) )\;.
\end{split}
 \end{equation}
Notice that, in the backward flow appearing in the last line, the first collision takes place at $t_1 \in (t-T,t)$, with rotation of the velocity of an angle $\theta_1$ determined by $n_1$; and after that collision (going backward) the trajectory performs a sequence of $k_1$ Larmor orbits, each one followed by a self-recollision with the same rotation $\theta_1$. 
 
 It is now enough to iterate Eq.\,\eqref{eq:Duhfm}. 
 
 In the first iteration, 
for any $t_1,n_1,k_1$ appearing in \eqref{eq:Duhfm} we write
 \begin{eqnarray} \label{eq:Duhfm'}
&& f^G(t_1-k_1T, \gamma_1^{-t+t_1-k_1T}(x,v)) \nonumber\\
&& =\chi(\{ t_1-k_1T<T\}) \,e^{-2(t_1-k_1T)} \, f_0(\gamma_0^{-t_1+k_1T}\gamma_1^{-t+t_1-k_1T}(x,v))	\nonumber\\
&&\quad+ 
\int_{(t_1-(k_1+1)T) \wedge 0}^{t_1-k_1T} dt_2 \cdots \;.
 \end{eqnarray}
Note that the constraint
 $$\chi(\{ 0<t_1-k_1T<T\})\chi(\{ (t-T) \wedge 0<t_1<t\}) \neq 0$$ implies that either $k_1 = [t/T]-1$ or $k_1=[t/T]$. Hence, when replaced into \eqref{eq:Duhfm}, the term in the second line of \eqref{eq:Duhfm'} 
produces a contribution 
\begin{align*}
& \int_{(t-T )\wedge 0}^{t} dt_1  \int dn_1 \left(n_1 \cdot \eta_0(t_1^+)\right)_+ e^{-2  t} \\
&\quad\cdot \sum_{k_1 = 0}^{[t/T]}   \chi(\{ 0<t_1-k_1T<T\}) f_0(\gamma_0^{-t_1+k_1T}\gamma_1^{-t+t_1-k_1T}(x,v))\\
&= e^{-2 t}\int_{(t-T) \wedge 0}^{t} dt_1 \int dn_1  \left(n_1 \cdot \eta_0(t_1^+)\right)_+
f_0( \gamma_1^{-t}(x,v))\;,
\end{align*}
where $\gamma_1^{-t}$ is the backward flow with first collision specified by  $\left(t_1, n_1\right)$, and with $[t/T]-1$ or $[t/T]$ self-recollisions.

Using the Boltzmann backward flow from $t_0 = t$ up to time zero and with $m=0,1,2$, the first iteration of Eq.\,\eqref{eq:Duhfm} can be written as follows:
 \begin{equation*}
\begin{split}
& f^G (t,x,v)=\chi(\{ t<T\}) \,e^{-2 t} \, f_0(\gamma_0^{-t}(x,v)) \\
&\;\;\;  + 
e^{-2 t}\int_{(t-T) \wedge 0}^{t} dt_1 \int dn_1 \left(n_1 \cdot \eta_1(t_1^+)\right)_+
f_0( \gamma_1^{-t}(x,v)) \\
& \;\;\; + 
\int_{(t-T) \wedge 0}^{t} dt_1 \int dn_1 \left(n_1 \cdot \eta_1(t_1^+)\right)_+
e^{-2 (t-t_1)} \sum_{k_1 = 0}^{[t/T]}  e^{-2 k_1 T}\\&
\;\;\;\;\;\;\cdot
\int_{(t_1-(k_1+1)T )\wedge 0}^{t_1-k_1T} dt_2 \int dn_2 \left(n_2 \cdot \eta_2(t_2^+)\right)_+
e^{-2 (t_1-k_1T-t_2)} \sum_{k_2=0}^{[(t_1-k_1T)/T]} e^{-2 k_2 T}\\&
\;\;\;\;\;\;\cdot
 f^G(t_2-k_2T,\gamma_2^{-t+t_2-k_2T}(x,v))\;.
\end{split}
 \end{equation*}

 Similarly, after the second iteration the sum over $k_2$ produces at most  two non-zero terms and the sum over $k_1$ can be performed:
 $$ \sum_{k_1 = 0}^{[t/T]}   \int_{(t_1-(k_1+1)T) \wedge 0}^{t_1-k_1T} dt_2 \,(\cdots)
 = \int_0^{t_1} dt_2\, (\cdots)\;,$$
which eliminates the constraints on $t_2$. 

The final formula is obtained after infinitely many iterations, thus concluding the proof. \qed
\bigskip
 
We conclude by constructing the Boltzmann process. We indicate by $ \sigma\left(\Pi_{t}\right)$ the Borel $\sigma$-field on $\Pi_{t}$ (cf.\,\eqref{eq:pathsp}-\eqref{eq:pathspm}).
For any $A \in \sigma\left(\Pi_{t}\right)$, we denote by $A_m$ the restriction of $A$ to $\Pi_{t,m}$. 
\begin{definition}[Boltzmann process] \label{def:BP}
Given $(x_0,v_0) \in \R^2 \times S^1$, the generalized Boltzmann process $t \to \zeta(t)$ starting from $(x_0,v_0)$ is the jump process in $D\left([0,t], \R^2\times S^1\right)$ defined by the path measure
 \begin{equation}
 \label{eq:GBP}
\begin{split}
P_t^{(x_0,v_0)} \left(A\right)&= e^{-2t}\chi(\{  t<T\})\,\delta_{A_0,\Pi_{t,0}} \ + e^{-2 t}\sum_{m \geq 1}\int_{A_m} \chi(\{  t_1 < T \})
\left[ \prod_{i=1}^m  \left(n_i \cdot \eta_m(t_i^-)\right)_- \right] \;,
\end{split}
 \end{equation}
$A \in \sigma\left(\Pi_{t}\right)$, where the (forward) Boltzmann flow \eqref{eq:GBF} in $[0,t]$ is computed with $t_0=0$, $(x,v) = (x_0,v_0)$
and $t_{m+1}=t$. We shall say that the ``obstacles'' $1,2,\cdots, m $ of the Boltzmann process are placed in $\xi_m(t_1),\cdots,\xi_m(t_m)$.
\end{definition}

It remains to make the link with Eq.\,\eqref{eq:Bmss}, which is expressed in terms of the backward (adjoint) flow. To do that, we write
$$
f^G (t,x,v) = \int_{\R^2\times S^1} dx'dv' \delta\left((x',v')-(x,v)\right)f^G (t,x',v')\;,
$$
insert \eqref{eq:Bmss} into the right hand side, and perform the change of variables:
$$
\left(x',v',t_1,n_1,\cdots,t_m,n_m\right) \to \left( \xi_m(0),\eta_m(0),t_1 - k_1T,R_{k_1\theta_1}(n_1),\cdots,t_m - k_mT,R_{k_m\theta_m}(n_m) \right)
$$
where $k_i = [\left(t_{i}-t_{i+1}\right)/T]$ and $\theta_i$ are the scattering angles in the backward Boltzmann flow. Notice that, by symmetry and by the scattering rule, 
$$\left(n_i \cdot \eta_m(t_i^+)\right)_+ = \left(R_{k_i\theta_i}(n_i) \cdot \eta_m(t_i^--k_iT)\right)_-\;.$$
The new variables are those of the forward Boltzmann flow. Renaming these integration variables
$\left(x_0,v_0,t_1,n_1,\cdots,t_m,n_m\right)$, and using \eqref{eq:GBP},
we get
\begin{equation} \label{eq:expproc}
f^G (t,x,v) = \int_{\R^2\times S^1}  dx_0\,dv_0 \, f_0(x_0,v_0) \sum_{m \geq 0}\int_{\Pi_{t,m}}dP_t^{(x_0,v_0)} 
\delta\left( \left(\xi_m(t),\eta_m(t)\right)  - (x,v)\right) \;.
\end{equation}

\section{Proof of Theorem \ref{thm:main}}\label{sec:proof}

The outline of the proof is standard. There are three main steps. We first eliminate suitable events of probability zero (Lemma \ref{lem:misurazero}); secondly, we parametrize the Lorentz process (cf. Definition \ref{def:flow}) in terms of impact times and impact vectors (Lemma \ref{eq:espansione}); and finally we show that, away from the excluded set, the differences with respect to the Boltzmann process (cf. Definition \ref{def:BP}) are negligible (Lemma \ref{lemma:shift-conv}). 

\begin{definition}[Pathologies]\label{def:patho}
Let $\left(\xi_m(s)\right)_{s \in [0,t]}$ be the generalized Boltzmann flow given as in Definition \ref{def:BFlow}. 
The pathological subset of $\Pi_t$ is given by
 $$
 {\cal N}_t = \bigcup_{m \geq 1} {\cal N}_{t,m}
$$
with
\begin{eqnarray*}
&&{\cal N}_{t,m} := \Big\{ \left(t_1,n_1,\cdots,t_m,n_m\right)\in\Pi_{t,m}\ \Big|\ 
t_1 < T\;,\ \mbox{$n_i\cdot\eta_m(t_{i}^-) \leq 0$ $\forall i$ and,}\\
&&\qquad\qquad
 \mbox{for some $j=1,\cdots,m$, $\left(\xi_m(s)\right)_{s \in [0,t]}$ crosses $\xi_m(t_j)$
more than $1+k_j$ times}\Big\}\;.
\end{eqnarray*}
\end{definition}
 Here we used \eqref{eq:defki}, and the convention $t_{m+1}=t$ to construct \eqref{eq:GBF}. 

\begin{lemma}\label{lem:misurazero}
{Let $P_t^{(x_0,v_0)}$ be the path measure for the generalized Boltzmann process defined in \eqref{eq:GBP}. We have }
$P_t^{(x_0,v_0)}\left({\cal N}_t\right) = 0\;.$
\end{lemma}

\smallskip
\noindent {\em Proof.}
Definition \ref{def:patho} implies that one of the following two events happens:

\smallskip
 \noindent {\bf (R)} the Boltzmann process returns to a collision point after having collided with a different obstacle (recollision):
 $$\inf_{s \in [t_{j+1},t]}|\xi_m(s) -\xi_m(t_{j}) | = 0\;;$$
 
 \smallskip
 \noindent {\bf (I)} the Boltzmann process encounters a new obstacle in a point already visited in the past (interference):
  $$\inf_{s \in [0,t_{j-1}+k_{j-1}T]}|\xi_m(s) -\xi_m(t_{j}) | = 0\;.$$

\smallskip
 
 Suppose that the infimum in (R) is attained first at time $\tau$. Either $t_{j+1}=t_j + k_jT$, or $\tau > t_{j+1}$. Assume the latter condition and that the last obstacle encountered before time $\tau$ is $j' \geq j+1$. Given $t_{j'}$ and a previous history, (R) can happen only for a finite number of values of $n_{j'}$. This follows from the fact that the self-recollision path around $j'$ is made of a finite number of orbits (which rotate rigidly with the impact vector). 
 
 Similarly (I) can happen only for a finite number of collision times $t_j$, with the following exception: $j \geq 2$ and the path in $[t_{j-1},t_j]$ is periodic. In the latter case, a continuous interval of collision times can realize the condition. A self-recollision path is periodic if and only if the scattering angle (hence $n_{j-1}$) is a rational fraction of $2\pi$. Namely, the magnetic field generates a dense set of periodic trajectories (of period $T, 2T, 3T, \cdots$), with repeated encounters with a single obstacle. On the other hand, such a periodic trajectory is completed only if its period is smaller than $t_j-t_{j-1}$. This is possible for $k_{j-1}$ values of the scattering angle. 
 
The above reasonings show that ${\cal N}_{t,m}$ has dimension lower than ${\Pi}_{t,m} $ and
$
| {\cal N}_{t,m} | = 0\;.
$
Since $P_t^{(x_0,v_0)} $ is absolutely continuous, the lemma is proved. \qed
\bigskip

By Lemma \ref{lem:misurazero}, it is enough to prove 
   \begin{equation} \label{eq:thm'}
 \lim_{\e \to 0}P_{\ep,t}^{(x_0,v_0)} = P_t^{(x_0,v_0)}\;.
 \end{equation}
on compact sets outside ${\cal N}_{t}$.

We turn now to the Lorentz process, and write down the explicit formula for the path measure. We shall assume $m \geq 1$ from now on, as in the case $m=0$ the proof of the theorem is straightforward. The following definition of Lorentz flow is similar in spirit to Definition \ref{def:BFlow}. The definition is such that, if the number of internal obstacles in $[0,t]$ is $m$, the Lorentz trajectory is given by the flow almost surely. 
\begin{definition}[Lorentz flow] \label{def:LF}
The Lorentz flow with $m$ {obstacles} and starting configuration $(x_0,v_0)$ is the Skorokhod trajectory
\begin{equation} \label{eq:LFm}
\zeta^\e_m(s) \equiv \left(\xi^\e_m(s),\eta^\e_m(s)\right)\;,\qquad s \in [0,t]
\end{equation}
constructed as follows.

Start from $\zeta^\e_m(0) = (x_0,v_0) \in \R^2 \times S^1$ for all $m$. 
We require $t_1 < T$ and apply the following rule, iteratively on $i=1,2,\cdots,m$ (with $t_0=0$):
\begin{itemize}
\item choose $\left(t_i,n_i \right)$ such that $t_{i-1} < t_i  <t$, and such that neither $\left(\zeta^\e_m(s)\right)_{s \in [0,t_{i-1}]}$, nor 
$$\zeta^\e_m(s) := \gamma_{\mbf{c}_{i-1},\e}^{s}\left( \left(x_0,v_0\right)\right)\qquad s \in [t_{i-1},t_i)\;,$$ 
intersect the open ball with center $\xi^{\e}(t_{i}) - \e n_i$; then set $c_i := \xi^{\e}(t_{i}) - \e n_i$, place a new obstacle in $c_i$ and apply the scattering $\zeta^\e_m(t_{i}^+) = S_{n_i}\left( \zeta^\e_m(t_{i}^-) \right)$ (cf.\,\eqref{eq:ode-flow'}).
 \end{itemize}
 The iteration is concluded in the last time interval by $\zeta^\e_m(s) := \gamma_{\mbf{c}_{m},\e}^{s}\left((x_0,v_0)\right)$ for $s \in [t_{m},t]$.
We call $\Pi^{\e, (x_0,v_0)}_{t,m}$ the set of admissible values $\left(t_1,n_1,\cdots,t_m,n_m\right)\in \Pi_{t,m}$, according to the iteration. 
\end{definition}
Recalling condition (I), we remark that 
\begin{equation} \label{eq:int}
\lim_{\e \to 0}
 \left\{ 0<t_1<\cdots t_m<t\,, t_1<T\;, n_i\cdot\eta_m(t_{i}^-) \leq 0\ \forall i\right\}
 \setminus \Pi^{\e, (x_0,v_0)}_{t,m} \subset {\cal N}_{t,m}\;. 
\end{equation}

\begin{lemma} \label{eq:espansione}
Let $A \in \sigma(\Pi_t)$ have non-empty component $A_m \in \Pi_{t,m}$ for only one value of $m \geq 1$.  Assume that $A$ is compact in $ \Pi_{t,m} \setminus {\cal N}_{t,m}$. 
 Then the path measure for the Lorentz process is given by
\begin{equation} \label{eq:Pete}
P_{\e,t}^{(x_0,v_0)}\left(A\right) =  \int_{A_m^\e} dt_1dn_1\cdots dt_mdn_m\,\chi(\{  t_1 < T \})\,\left[ \prod_{i=1}^m  \left(n_i \cdot \eta^\e_m(t_i^-)\right)_- \right] \,\frac{e^{-\e^{-1}|\cal{T}^\e|}}{1-\pi\e}
\end{equation}
where
\begin{equation} \label{eq:int'}
A_m^\e \equiv  A_m^{\e,t,(x_0,v_0)}:= A_m \cap \Pi^{\e, (x_0,v_0)}_{t,m}
\end{equation}
and
$$
 {\cal{T}}^\e \equiv {\cal{T}}^{\e,t,(x_0,v_0)}_{m}
:= \left\{ x \in \R^{2}\ \Big|\ 
\inf_{s \in [0,t]}|x - \xi^\e_m(s)| \leq \e
\right\}\;.
$$
\end{lemma}

\bigskip
\noindent {\em Proof.}
Using the notation introduced in \eqref{eq:zet}, the Lorentz process is
$$\zeta^\e(s) = \left(\xi^\e(s),\eta^\e(s)\right)=\gamma_{\mbf{c}_N,\e}^{s}(x_0,v_0)\;,\qquad s \in [0,t]\;.$$
Since we restrict to the $m-$obstacle path space, if we rename $c_1,\cdots,c_m$ the centers of the internal obstacles in $[0,t]$ (cf.\,Def.\,\ref{def:int-ext}), we can write 
$$\zeta^\e(s) = \gamma_{\mbf{c}_m,\e}^{s}(x_0,v_0)\;,\qquad s \in [0,t]\;.$$
These internal obstacles can be chosen in $N(N-1)\cdots (N-m+1)$ ways.

If $\mathcal{A}\subset\R^2$ is any bounded measurable set such that
the Lorentz process stays inside $\mathcal{A}$ in the time-interval $[0,t]$, we then have that
\begin{eqnarray*}
&&P_{\e,t}^{(x_0,v_0)}\left(A\right) =  \frac{e^{-\e^{-1} |\mathcal{A}|}}{1-\pi\e} \e^{-m}\int_{\mathcal{A}^{m}} d\mbf{c}_{m}
\,\chi\left(\left\{\left(t_1,n_1,\cdots,t_m,n_m\right)\in A_m\right\}\right) \chi(\{  t_1 < T \})\\
&&\qquad\qquad\qquad \cdot
\sum_{m' \geq 0} \frac{\e^{-m'}}{m'!} \int_{\mathcal{A}^{m'}}d\mbf{c}'_{m'}
\prod_{i=1}^{m'} \chi\left(\left\{ \inf_{s \in [0,t]}|c'_i - \xi^\e(s)| > \e \right\}\right)\;,
\end{eqnarray*}
where $\left(t_1,n_1,\cdots,t_m,n_m\right)$ is the collection of impact times and impact vectors according to Definition \ref{def:times-vectors}. Notice that the right hand side of \eqref{poisson} has been used with $\mu = \e^{-1}$ and renormalized by a factor $(1-\pi\e)^{-1}$, in such a way to ensure $|c_i-x_0| \geq \e$ for all $i$. 

In a more compact form,
\begin{equation}\label{eq:Pecf}
P_{\e,t}^{(x_0,v_0)}\left(A\right) =  \frac{\e^{-m}}{1-\pi\e}\int_{\tilde A_m^\e} d\mbf{c}_{m} \,\chi(\{  t_1 < T \})\,e^{-\e^{-1}| \tilde{\cal{T}}^\e|}
\end{equation}
with 
\begin{eqnarray*}
&& \tilde A_m^\e \equiv \tilde A^{\e,t,(x_0,v_0)} := \left\{ \mbf{c}_{m} \in \mathcal{A}^{m}\ \Big|\ 
\left(t_1,n_1,\cdots,t_m,n_m\right)\in A_m
\right\}\;,\\
&& \tilde{\cal{T}}^\e \equiv \tilde{\cal{T}}^{\e,t,(x_0,v_0)}_{\mbf{c}_{m}}
:= \left\{ x \in \R^{2}\ \Big|\ 
\inf_{s \in [0,t]}|x - \xi^\e(s)| \leq \e
\right\}\;.
\end{eqnarray*}

Note now that, by construction, we can replace $A_m$ in the above formulas with $A_m^\e \subset A_m$. Indeed $A_m^\e$ is defined as the subset for which there actually exists a mechanical trajectory with $m$ obstacles, having impact times and impact vectors in $A_m$. This forbids values of $(t_i,n_i)$ such that the obstacle centered in $c_i$ would overlap with points already visited in the past. 

We are ready to perform the change of variables
$$
\tilde A_m^\e \ni \left(c_1,\cdots, c_m\right) \longrightarrow \left(t_1,n_1,\cdots,t_m,n_m\right) \in A_m^\e
$$
inside \eqref{eq:Pecf}.
A standard computation of the Jacobian determinant leads then  to \eqref{eq:Pete}. \qed
\bigskip

Next we discuss the convergence of \eqref{eq:LFm} to \eqref{eq:GBF}. 
\begin{lemma}\label{lemma:shift-conv}
Let $t>0$ and $m \geq 1$. Let $A_m \in \Pi_{t,m} \setminus {\cal N}_{t,m}$ be a compact set.
Then there exists a constant $C>0$ such that, if $\e$ is small enough, 
for any admissible path $(t_1,n_1,\cdots,t_m,n_m)$ in $A_m$, 
the Lorentz flow does not allow recollisions and
\begin{equation*}\label{eq:shift-conv*}
\sup_{s \in [0,t]} |\xi^\e_m(s)-\xi_m(s)| \leq C\left[t/T\right] \e\;,\qquad
\max_{i=1,\cdots,m} |\eta^\e_m(t^-_i)-\eta_m(t^-_i)| \leq C\left[t/T\right] \e\;.
\end{equation*}
\end{lemma}
{Notice that the estimates are uniform in $m$ and that $\left[t/T\right]$ is a bound on the maximal number of self-recollisions.}

\bigskip
\noindent 
{\em Proof.}
We introduce a stopping time $\tau^\e$, defined as the first time such that the Lorentz process has a recollision which is not a self-recollision (cf.\,Def.\,\ref{def:recollision}).
By definition, $\tau^\e > t_1$ and $\zeta^\e_m(s)=\zeta_m(s)$ for $s \in [0,t_1]$.
Moreover for $s \in [0,\tau^\e)$, the differences between $\zeta^\e_m(s)$ and $\zeta_m(s)$ are exclusively due to the self-recollisions, which we discuss next.

Each self-recollision orbit in the Lorentz flow differs of an error $O(\e)$ from the corresponding orbit of the Boltzmann flow, according to \eqref{eq:err}. 
Notice that also the self-recollision times differ, in the two flows, of an amount $O(\e)$ (the Lorentz process jumps to a new orbit slightly before the Boltzmann process). During this discrepancy of times, a big difference between the velocities is generated. What is important is that, for $\e$ small, the time discrepancies in self-recollisions with the obstacle $i$ are all smaller than $(k_i+1)T-(t_{i+1}-t_i)$: thus the total number of self-recollisions in the Lorentz process is $k_i$, exactly as in the Boltzmann flow. 
Since the total number of collisions is finite, it follows that: 

\smallskip
\noindent
{\bf (a)} the difference between $\eta^\e_m(t_i^-)$ and $\eta_m(t_i^-)$ is $O(\e)$ for all $i$ with $t_i < \tau^\e$; 

\smallskip
\noindent
{\bf (b)} the distance between $\xi^\e_m(s)$ and $\xi_m(s)$ is $O(\e)$ for $s \in [0,\tau^\e)$. In particular, $c_j = \xi^\e_{m}(t_j)-\e n_j$ is $O(\e)$ away from $\xi_{m}(t_j)$ if $t_j < \tau^\e$.

\smallskip
As $A_m$ is compact outside ${\cal N}_{t,m}$,  the Boltzmann process has the following property:
 there exists a $\delta>0$ such that each arc of Larmor orbit keeps a distance larger than $\delta$ from the obstacles which are not at the extremal points of the arc. In formulas:
\begin{eqnarray*}
&& \min_{j= 2,\cdots,m}\,\inf_{s \in[0,t_1]} |\xi_m(s)-\xi_{m}(t_j)|> \delta\;,\nonumber \\
&&
\min_{\substack{i,j = 1,\cdots,m \\ j \neq i}}\,\inf_{s \in [t_i,t_i+k_iT]} |\xi_m(s)-\xi_{m}(t_j)|> \delta\;,\nonumber \\
&& 
\min_{\substack{i,j = 1,\cdots,m \\ j \neq i,i+1}}\,\inf_{s \in [t_i+k_iT,t_{i+1}]} |\xi_m(s)-\xi_{m}(t_j)|
> \delta\;. 
\end{eqnarray*}
This and the above remark (b) on self-recollisions imply that, for $\e$ small enough, 
$$\min_{\substack{i=1,\cdots,m \\ j = 1,\cdots, i-1}}\inf_{s \in [t_{i},t]}|\xi^\e_m(s) -c_j | > \delta/2 > \e\;.$$
Which proves $\tau^\e > t$. \qed
\bigskip

\noindent
{\em End of the proof of Theorem \ref{thm:main}.}
Our purpose is to compare formulas \eqref{eq:Pete} and \eqref{eq:GBP}. 
Remind that the sums over $m$ are absolutely convergent (uniformly in $\e$) for arbitrary sequences $A_m$. Therefore it is enough to prove convergence of \eqref{eq:Pete} to the $m$-th term in \eqref{eq:GBP}.

Observe that, under the assumptions of Lemma \ref{eq:shift-conv*}, the region ${\cal{T}}^\e$ reduces to a union of $m$ daisies, plus a first stem:
$$
{\cal{T}}^\e = \bigcup_{i=0}^{m} \mathcal{D}^\e_{i}\;,
$$
having used Definition \ref{def:daisy} with $t_{m+1}=t$. 

Since $|\mathcal{D}^\e_{i}| \leq 2\e \left( t_{i+1}-t_i\right)$
we deduce from Lemma \ref{lem:misurazero}, \eqref{eq:int}, \eqref{eq:int'}, \eqref{eq:Pete}, Lemma \ref{lemma:shift-conv} and \eqref{eq:GBP} that
\begin{equation*}
\lim_{\e \to 0} P_{\e,t}^{(x_0,v_0)}\left(A\right) \geq \lim_{\e \to 0} e^{-2t}\,  \int_{A_m^\e} \,\chi(\{  t_1 < T \})\,\left[ \prod_{i=1}^m  \left(n_i \cdot \eta^\e_m(t_i^-)\right)_- \right]  = P_{t}^{(x_0,v_0)}\left(A\right)\;,
\end{equation*}
by dominated convergence. Using the normalization condition
$$
P_{\e,t}^{(x_0,v_0)}\left(\Pi_{t}\right) = P_{t}^{(x_0,v_0)}\left(\Pi_{t}\right) = 1-e^{-2T} \delta_{t>T}\;,
$$
and Lemma \ref{lem:misurazero}, we finally obtain Eq.\,\eqref{eq:thm}.
Eq.\,\eqref{thm:cor} follows from \eqref{eq:expect-density} and \eqref{eq:expproc}.
\qed

\bigskip

\noindent
{\bf Acknowledgements.} 
The authors are grateful to A.V. Bobylev who brought this problem to their attention.  
A.N. acknowledges support through the CRC 1060 \emph{The mathematics of emergent effects} of the University of Bonn that is funded through the German Science Foundation (DFG).
C.S. acknowledges the support of the SNSF through the project  PCEFP2\_181153 and of the NCCR SwissMAP.\\
The hospitality of the MFO during the mini-workshop 1910b ``Lorentz Gas Dynamics'' and of the HIM during the Junior Trimester Program ``Kinetic Theory'' in 2019 are gratefully acknowledged.  \\
{We thank the anonymous referee for comments that helped improving the manuscript.}

\bigskip
\adresse

\end{document}